\documentclass{aa}  

\usepackage{graphicx}
\usepackage{txfonts}
\usepackage{hyperref}
\defcitealias{Davoult2024}{D24}
\usepackage{natbib}

\begin{document}

   \title{Earth-like planet predictor: a machine learning approach}

   \author{Jeanne Davoult
          \inst{1,2,3}
          \and
          Romain Eltschinger
          \inst{1,2}
          \and
          Yann Alibert\inst{4}
          }

   \institute{Space research \& Planetary Sciences (WP), Universität Bern, Gesellschaftsstrasse 6, 3012 Bern, Switzerland
              \and
              NCCR PlanetS, Universität Bern, Gesellschaftstrasse 6, 3012 Bern, Switzerland
              \and
              Institute of Planetary Research, German Aerospace Center, Rutherfordstrasse 2, 12489 Berlin, Germany\\
              \email{jeanne.davoult@dlr.de}
              \and
             Center for Space and Habitability (CSH), Universität Bern, Gesellschaftstrasse 6, 3012 Bern, Switzerland
             }

   \date{Received 01/10/2024; accepted 28/02/2025}

% \abstract{}{}{}{}{}
% 5 {} token are mandatory
 
  \abstract
   {Searching for planets analogous to Earth in terms of mass and equilibrium temperature is currently the first step in the quest for habitable conditions outside our Solar System and, ultimately, the search for life in the universe. Future missions such as PLAnetary Transits and Oscillations of stars (PLATO) or Large Interferometer For Exoplanets (LIFE) will begin to detect and characterise these small, cold planets, dedicating significant observation time to them. }
   {The aim of this work is to predict which stars are most likely to host an Earth-like planet (ELP) to avoid blind searches, minimises detection times, and thus maximises the number of detections.}
   {Using a previous study on correlations between the presence of an ELP and the properties of its system, we trained a Random Forest to recognise and classify systems as `hosting an ELP' or `not hosting an ELP'. The Random Forest was trained and tested on populations of synthetic planetary systems derived from the Bern model, and then applied to real observed systems.}
   {The tests conducted on the machine learning (ML) model yield precision scores of up to 0.99, indicating that 99\% of the systems identified by the model as having ELPs possess at least one. Among the few real observed systems that have been tested, eight have been selected as having a high probability of hosting an ELP, and a quick study of the stability of these systems confirms that the presence of an Earth-like planet within them would leave them stable.}
   {The excellent results obtained from the tests conducted on the ML model demonstrate its ability to recognise the typical architectures of systems with or without ELPs within populations derived from the Bern model. If we assume that the Bern model adequately describes the architecture of real systems, then such a tool can prove indispensable in the search for Earth-like planets. A similar approach could be applied to other planetary system formation models to validate those predictions.}


   \maketitle
%
%________________________________________________________________

\section{Introduction}
Detecting planets as small and cold as Earth is a major technical challenge in exoplanet research for the coming decades. The upcoming PLAnetary Transits and Oscillations of stars mission \citep[PLATO;][]{Rauer2014} and the concept of mission Large Interferometer For Exoplanets \citep[LIFE;][]{Kammerer2018, LIFE1} will be dedicated to this task, but their long periods (potentially 1~year or more) consume significant observation time. Although various studies on planet demographics suggest that small terrestrial planets with short periods are very common around main sequence stars \citep[e.g.][]{Mayor2011,Tuomi2019,Kunimoto&Matthews2020}, the abundance of terrestrial planets with longer periods in the habitable zone of their star is more uncertain \citep[e.g.][]{Hsu2019,Bryson2021}. Understanding and anticipating where Earth-like planets (ELPs in the rest of the paper) form first, and thus targeting observations to avoid blind searches, minimizes the average observation time for detecting an ELP and maximizes the number of detections. Studies conducted on the architecture and correlations in multi-planet systems over the years \citep[e.g.][]{Lissauer2011,Millholland2017,Weiss2018,Gilbert2020,Mishra2023,Emsenhuber2023,Davoult2024} among others)  have highlighted correlations between the properties of planets in the same system. For example, correlations have been discovered between the presence of an inner terrestrial planet and the presence of an outer giant planet \citep[e.g.][]{Zhu&Wu2018,Zhu2024,Bryan&Lee2024}, but it exists an anti-correlation between the presence of a hot Jupiter and the `peas-in-a-pod' formation \citep{Weiss2018, Latham2011, Steffen2012}. Thus, the architecture of systems, representing the arrangement of planets in a system, is not the result of chance but of simultaneous formation within the same system. In other words, the planets in the same system bear the imprint of each other's formation. Therefore, detected planets could provide insights into undetected planets within the same system.\\
Attempts to predict yet-undetected exoplanets based on detected exoplanets' properties have emerged in recent years. For example, \cite{Bovaird2013}, \cite{Bovaird2015}, \cite{Lara2020} and \cite{Mousavi-Sadr2021} have attempted to use a logarithmic relationship between planetary periods, akin to the Titius-Bode law, to predict missing planets within systems. Similarly, \cite{Dietrich2020} and \cite{Sandford2021} utilised statistical data from already-detected planetary populations to forecast future observations. However, all these previous studies relied on data from observed exoplanet populations. Here, we propose using synthetic planetary systems from the Bern model --- systems in which all planets are known --- avoiding observational bias.\\
In a previous study (\cite{Davoult2024}, \citetalias{Davoult2024} in the rest of the paper), we have established correlations between the presence of an ELP in the temperate zone of its star and other properties of its system, including the architecture of the planetary system as described in the paper, and the mass, radius, and period of the innermost detectable planet (IDP in the rest of the paper) of that system. In this study, we present the results of algorithms using a machine learning (ML) model capable of learning the differences in properties between systems hosting an ELP and those not hosting an ELP, in order to predict whether a given system hosts an ELP or not.\\
The use of ML models requires very large datasets, which makes it impossible to use only data from observed systems. In addition to the small number of known planetary systems to date (just under 5000 in July 2024), there is the problem of partial knowledge of these systems. ELPs, being small and relatively cold planets, are difficult to detect using the most efficient detection methods (i.e., transits and radial velocities). Indeed, only 24 systems with at least one ELP are known (following the definition of Sect. \ref{sec:ELP}), representing 0.5\% of all systems observed to date. Herefore, using those data in a ML-based approach is impossible.\\
To address these two major problems, this study utilises populations of several thousand synthetic planetary systems generated from the Bern model. Studies have examined the outputs of this model and compared them to observed systems \citep[e.g.][]{Mulders2019, NGPPS3,NGPPS4,NGPPS6,Mishra2023,Davoult2024,Emsenhuber_inprep}, revealing that these synthetic systems possess similar system-level characteristics as observed systems---such as similar architectures \citep{Mishra2023, Davoult2024}, recurring patterns in Peas-in-a-Pod \citep{NGPPS6}, correlations between outer giants and inner earths \citep{NGPPS3}, etc. These comparisons lead us to believe that synthetic systems generated from the Bern model serve as reasonable training data for ML models. Additionally, a study \citep{NGPPS5} using a data-driven approach was also successfully conducted with the synthetic planetary system populations from the Bern model, aiming to predict the types of planets in a system based on the initial conditions of the protoplanetary disk and planetary embryos.\\
Section \ref{sec:BernModel} briefly describes the Bern model and the populations used. Section \ref{sec:method} outlines the various ML models, observational biases, and system features used. In Section \ref{sec:results}, we describe the results obtained for the different models, and we discuss and conclude in Section \ref{sec:conclusion}.

%_______________________________________________________________________________
\section{Synthetic population of planetary systems}\label{sec:BernModel}
%_______________________________________________________________________________

%-------------------------------------------------------------------------------
   \subsection{The Bern model and synthetic populations}
%-------------------------------------------------------------------------------

The planetary system formation and evolution model used in this study is the Generation III of the Bern model, described in detail in \cite{NGPPS1}. This global model utilises the population synthesis method, as explained in detail in \cite{Mordasini2018}, and is based on the core accretion paradigm \citep{Pollack1996}. The planetary formation is modelled over 20~Myr, during which 20 planetary embryos embedded in a disk of gas and planetesimals accrete material to form planets, migrate, and dynamically interact, leading to ejections, giant impacts, or resonance traps.
At the end of this formation phase, the model tracks the planets' thermodynamical evolution (consisting mainly of cooling and contraction) for 10~Gyr. During this evolution phase, atmospheric escape and tidal migration are also monitored. For more details on the parameterisation of the protoplanetary disk and the various physical processes involved in the formation and evolution of planets, refer to \cite{NGPPS1,NGPPS2}.\\
In a population synthesis, some parameters are fixed while others vary. In the populations of planetary systems used in this study, the fixed general parameters of the systems include the mass of the central star (1, 0.5, or 0.2~$M_{\odot}$), the number of planetary embryos (20), the gas viscosity ($\alpha$ = 2 $\times$10$^{-3}$), the distribution of gas and planetesimals in the protoplanetary disk \citep{Veras2004}, the size of the planetesimals (radius = 300~m), and their density (rocky 3.2 g cm$^{-3}$, icy 1g cm$^{-3}$). The rest of the initial conditions are randomly drawn according to a probability distribution constrained by observations, which allows for diversity in the resulting synthetic planetary systems. The variable parameters include the initial mass of the gas disk, $M_g$ \citep{Beckwith+1996}, the external photo-evaporation rate $M_{\text{wind}}$ \citep{Haisch2001}, the dust-to-gas ratio, f$_{\text{D/G}}$=$M_\text{s}$/$M_\text{g}$ (where $M_s$ is the mass of the solid disk) \citep{Murray2001, Santos2003}, the inner edge of the gas disc, $R_{\text{in}}$, and the initial location of the embryos.\\
   The three populations of synthetic systems used in this study differ only in the mass of the central star. This single difference directly influences the mass of the protoplanetary disk and thus the amount of material available for planet formation. As a result, the three populations exhibit different occurrences and properties for the same type of planet, highlighting the importance of studying various types of stars.\\
   The three populations used are:
   \begin{itemize}
       \item G-pop: 24 365 systems around solar mass stars
       \item earlyM-pop: 14 559 systems around 0.5 solar mass stars
       \item lateM-pop: 14 958 systems around 0.2 solar mass stars.
   \end{itemize}
   For a detailed analysis of the different types of planets and their occurrences in the above populations, refer to \citetalias{Davoult2024}.

   %----------------------------------------------------------------------------
   \subsection{Earth-like Planet} \label{sec:ELP}
   %----------------------------------------------------------------------------
    This study aims to predict which systems host an Earth-like planet or not. The ELP category refers to a small terrestrial planet with a mass ranging from 0.5 to 3~$M_{\oplus}$, orbiting the temperate zone of its star. The mass range was chosen in accordance with the work of \cite{Kopparapu2018} and \cite{NGPPS4}. The temperate zone, defined in \cite{Davoult2024}, is defined much broader as the habitable zone and extends in terms of equilibrium temperature ($T_{\text{eq}}$) from 160 to 510~K, calculated as follows:
   \begin{equation}
   	T_{\text{eq}}[\text{K}] = 279 \cdot a[\text{AU}]^{-1/2}\cdot L_{\star}[\text{L}_{\odot}]^{1/4},
   \end{equation} 
   where $a$ is the semi-major axis of the planet and $L_{\star}$ is the luminosity of the star.
   This correspond to a zone between 0.39 to 3.9 AU around a G-type star, between 0.25 to 2.52 around a early-M type star and between 0.15 to 1.48 around a late-M type star. By extending the target zone, we increase the number of systems with an ELP, and we reduce the imbalance in terms of proportion in the data, which is beneficial for ML models.\\ 
   As seen in \citetalias{Davoult2024}, the occurrence of a certain type of planet varies depending on the type of star it orbits. Thus, in our three populations, we find 60\% of systems with an ELP around solar-mass stars, 74\% around stars of 0.5~$M_{\odot}$, and 40\% around stars of 0.2~$M_{\odot}$.

   \subsection{Correlations between ELP and the properties of their systems}
   In \citetalias{Davoult2024}, we investigate correlations between planets in the synthetic planetary systems from the Bern model and their architecture to define a typical profile of a system hosting an ELP. Our conclusions highlight a correlation between the presence of an ELP, the architecture of its system, and the properties of the innermost detectable planet (IDP). Indeed, Earth-like planets tend to form in systems mainly composed of low-mass planets (M < 20 $M_{\oplus}$). In systems with more massive planets, the properties of the IDP, such as mass, radius, and period, can be indicators of ELP presence. A small, low-mass IDP suggests in-situ formation in a low-mass disk, while a giant IDP suggest a massive disk and/or planetary migration, unfavorable for a stable Earth-like planet in the habitable zone. The IDP's period indicates the positions of other planets: a close-in IDP suggests inward planet grouping, leaving the HZ empty, while slightly longer periods (> tens of days) indicate outward grouping, increasing HZ planet probability. Thus, ELP presence correlates with the system's architecture, and IDP's mass, radius, and period.\\
   Table 7 of \citetalias{Davoult2024} summarises the conditional probabilities of ELP's presence in systems according to the mass of the central star, the observed system architecture, and the properties of the IDP, providing an overview of the combinations most favourable for ELP formation in a system.
   The present paper uses part of their results to develop a predictor incorporating a ML model that the community can use to predict whether a system is likely to host an ELP or not based on its observable properties. We relied on the work presented in \citetalias{Davoult2024} to define the observable properties used in this prediction.
   
\section{Method} \label{sec:method}
\subsection{Machine learning classifier (MLC)}\label{sec:ML}
In ML methodologies, algorithms typically perform two main tasks: classification and regression. This problem is a case of classification, aiming to classify a system into the categories of `hosts an ELP' or `does not host an ELP'. ML models are trained to recognise data falling into one category or another using a dataset of thousands of data points. Once trained, they can predict, on an independent dataset, which class an instance falls into.\\
There are many classifier tools available, with the most common being decision trees, support vector machines (SVMs), or Random Forests, among others. Random Forests fall into the category of `ensemble' learning methods. They consist of multiple sub-classifiers, with each `local' classifier trained on a subset of data (which is not the entire training set). Then, all local classifiers are queried to classify an element. The `global' classifier (which includes all local classifiers) decides based on the majority of vote: if a majority of local classifiers vote to classify an element into the `True' category, then the final response of the global classifier is `True'. In this case, the category `hosts an ELP' is True, and the category `does not host an ELP' is False.\\
We aim to predict whether a system hosts an ELP or not to target observations to avoid wasting observation time. Therefore, we want to ensure that the positive responses given by our algorithm can be trusted, meaning it produces very few `false positives'. To ensure this, we want to maximise the precision score (PS), which measures the ratio of `true positives' to all elements labelled as `positive' (true positives and false positives). The precision score is the ability of the classifier not to label a negative sample as positive:
\begin{equation}
    \text{PS} = \frac{\text{True Positive}}{\text{True Positive} + \text{False Positive}}.
\end{equation}
When the precision score increases, the recall score (RC) decreases. The recall score is the ability of the classifier to find all the positive samples and is computed as follows:
\begin{equation}
    \text{RC} = \frac{\text{True Positive}}{\text{True Positive} + \text{False Negative}}.
\end{equation}
In other words, the more we focus on elements most likely to be labelled `True', the more we miss true positives in the batch. This is not necessarily a major issue because we do not particularly want to maximise the RC. The False Negative rate is characterised by the RC. The lower the RC, the higher the number of false negatives. Given the time required to detect an Earth-like planet, we chose to focus on maximising the PS rather than the RC. It seems more important to ensure the method's reliability by concentrating on the most robust systems. Indeed, there exist many potential targets for a limiting telescope time. In the opposite situation (plenty of telescope time, but few targets), we would like to optimise the recall score (minimising the number of false negatives). The issue arises when the classifier fails to find any positives in our efforts to maximize true positives.\\
The significant advantage of ensemble algorithms is adjusting the voting threshold ourselves. Instead of declaring the threshold at 50\% of local classifiers as the threshold for the global classifier to decide, we can adjust this threshold. In the rest of the study, we examine several thresholds: 70\%, 80\%, and 90\%. We define `voting rate' as the proportion of local classifiers that have classified an instance as True.\\
The Random Forest used in this study is made up of 500 decision trees, allowing to reduce the variance through ensemble learning, while keeping the training time reasonable. Each Decision Tree is trained on a minimum sample of 100 instances in order to increase the diversity between the trees, while allowing the generalisation of the classification. Trees trained on fewer instances have a tendency to learn details and overfit. Finally, the maximum depth of each Tree is limited to five in order to limit the complexity of the model, forcing it to capture only the most important relationships in the data. Tests conducted with a higher maximum depth (no limitation) did not change the results very much, proving the stability of the results.
\subsection{Observational bias}\label{sec:bias}
To use the observable properties of synthetic planetary systems from the Bern model, we apply an observational bias to retain only the planets that could be theoretically observed. This bias involves a radial velocity (RV) semi-amplitude threshold on the star. Planets with an RV semi-amplitude above this threshold are considered detected, while those below are considered undetected. The detected planets form the new planetary system from which we extract the characteristics used by the ML model.
The RV semi-amplitude that a planet induces on its star is calculated as follows:
\begin{equation}
K_{\text{RV}}[\text{m}\cdot \text{s}^{-1}] = 0.6395 \cdot P[\text{days}]^{1/3}\cdot M_p[\text{M}_{\oplus}]\cdot M_{\star}[\text{M}_{\odot}]^{-2/3},
\end{equation}
Where $P$ is the period of the planet, $M_p$ its mass and $M_{\star}$ the mass of the star.\\
The detection threshold is set to exclude ELPs from this study. Ignoring systems with detected ELPs is reasonable, given that only 24 systems (0.5\%) among nearly 4900 observed (as of July 2024) are known to host a planet following our definition (see Sect. \ref{sec:ELP}). 
The RV threshold for detectability varies between populations due to two factors: the limits of the temperate zone vary depending the population ($T_{\text{eff}} \propto L_{\star}^{\text{1/4}}$) and for a given planetary mass and period, the RV semi-amplitude signal varies as $M_{\star}^{\text{-2/3}}$. The values used are presented in Table \ref{tab:threshold}.

   \begin{table}[h]
   	\caption[]{RV semi-amplitude thresholds retained for each population.}
   	\begin{tabular}{cr}
   		\hline
   		Central star's mass & threshold retained \\
   		1~$M_{\odot}$ (G-pop)   & 0.43 m$\cdot$s$^{-1}$\\
   		0.5~$M_{\odot}$ (earlyM-pop) & 0.76 m$\cdot$s$^{-1}$\\
   		0.2~$M_{\odot}$ (lateM-pop) & 1.55 m$\cdot$s$^{-1}$\\
   		\hline
   		\label{tab:threshold}
   	\end{tabular}
   \end{table}
Although this observation bias is too simple to be considered accurate, \citetalias{Davoult2024} have shown that it can reproduce the proportions in architectures observed in multiplanet systems, which is sufficient for this study. An analysis of the impact of this bias on the synthetic populations of planetary systems used here is available in \citetalias{Davoult2024}.

\subsection{Features of interest} \label{subsec:features}
When using a ML model, it projects the dataset into an N-dimensional space, where N is the number of dimensions of the dataset. In our case, N represents the amount of information about each system provided to the algorithm for learning. Each of the three populations contains between $\sim$15 000 and $\sim$25 000 systems. After removing empty systems (systems with no planets) and systems with no visible planets (systems with planets but that cannot be classified in our architecture classes), only about 5 000 to 20 000 instances remain in each population. While the size of this dataset allows us to conduct this study, it remains limited. If N is too large, the data may become lost in a high-dimensional space, making the task challenging for the model and increasing the risk of overfitting. Therefore, it is important to describe each instance --- each system in this case --- with a reasonable number of features to mitigate the risk of overfitting. The challenge lies in selecting the right features, the most useful ones that provide the most information. Given that the aim of this project is observational, the information provided to the ML model must be easily observable quantities. We present two strategies: the first strategy utilises the findings of \citetalias{Davoult2024}, while the second strategy involves defining the features based on a manual analysis.

\subsubsection{Observables derived from \citetalias{Davoult2024}} \label{subsubsec:featuresD24}
In \citetalias{Davoult2024} we present a study of correlations between the presence of an ELP in a system and observable quantities of those systems. The conclusions link the presence of an ELP with a system's `biased' architecture, as well as the mass, radius, and period of the innermost detectable planet (IDP). The biased architecture of a system refers to the architecture of a system considering only the detectable planets in that system. The method used to calculate the observational bias is the same in this article as in \citetalias{Davoult2024}, ensuring a similar approach.\\
In \citetalias{Davoult2024}, we also introduce a method for classifying each system into a different architecture class using Principal Component Analysis (PCA) applied in the mass--semi-major axis plane of the visible planets in the system, along with the mass of the most massive visible planet in the system and the number of visible planets. Thus we define five classes :
\begin{itemize}
    \item Low-mass: systems with at least two visible planets in which all planets are less massive than 20~$M_{\oplus}$\\
    \item Anti-Ordered: systems with at least two visible planets, with at least one planet more massive than 20~$M_{\oplus}$ and a general tendency for the planetary masses to decrease with the distance to star increasing.\\
    \item Ordered: systems with at least two visible planets, with at least one planet more massive than 20~$M_{\oplus}$ and a general tendency for the planetary masses to increase with the distance to the star increasing.\\
    \item Mixed: systems with at least two visible planet and a planet more massive than 20~$M_{\oplus}$, and a large variability in the planetary masses, inducing no special tendency.\\
    \item n~=~1: systems with only one visible planet
\end{itemize}
These four descriptive features make up the first set: the architecture of the visible system, and the IDP's mass, radius, and period.

\subsubsection{Manual feature selection} \label{subsubsec:featuresmanual}
Looking at the systems generated from the Bern model, it is evident that systems with ELPs are very similar to each other, whereas, conversely, they are very different from systems without ELPs.
\begin{figure*}[h]
    \centering
    \includegraphics[width=9cm]{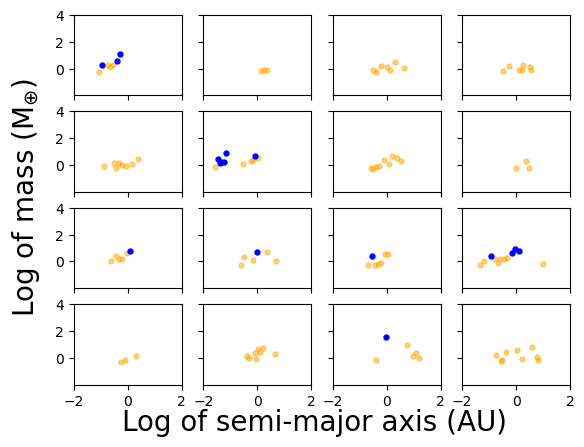}
    \includegraphics[width=9cm]{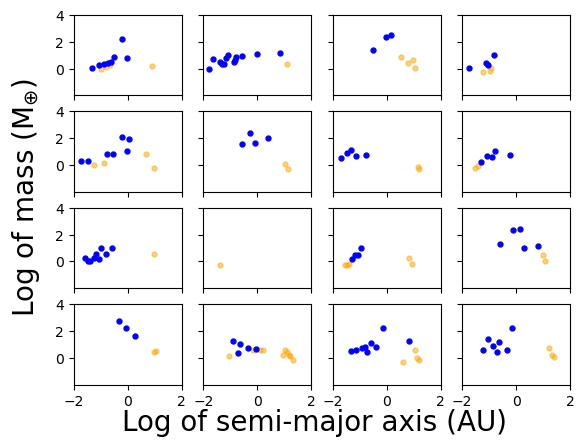}
    \caption{Representation of 16 systems with ELP (left) and 16 systems without ELP (right) in a semi-major axis - planetary mass diagram (in log scale for both axes). Blue dots represent `detectable' planets and yellow dots `undetectable' planets.}
    \label{fig:comparison_syst}
\end{figure*}
Figure \ref{fig:comparison_syst} depicts two types of systems: on the left are sixteen systems with at least one ELP randomly selected from the Sun-like stars population, and on the right are sixteen systems without ELPs randomly selected from the same population. The blue dots represent planets that have passed the detection threshold, while the yellow dots represent `non-detectable' planets. Systems without ELPs (left) are all very similar to each other. They consist mostly of small planets, with relatively few detectable planets, and few planets more massive than Neptune. Additionally, compact, clustered systems are observed around one AU. In contrast, systems without ELPs, on the right, display more detectable planets, including more massive planets. It is common to find a giant or at least a sub-giant planet in these systems. The systems are more spread out in terms of semi-major axis range, but we can still find clusters of small terrestrial planets, which are shifted inward, very close to the star, at a fraction of an AU. These visible differences allow us to easily classify a system as `host an ELP' or `does not host an ELP'. To use these features in an ML model, we need to quantify them, describing each system with a limited number of features.\\
Our choice of features, which we believe best capture the differences observed visually, is as follows:
\begin{itemize}
    \item number of visible planets 
    \item number of giant planets ($M_p > 100~M_{\oplus}$)
    \item IDP's mass,
\end{itemize}
to which we add the star's mass, known to be correlated with the type of planets present in the system.
Indeed, as studied in Section \ref{sec:ELP}, the proportion of systems with ELPs is not the same in the three populations because the central star's mass plays a role in planetary formation. These five features make up our second set.

\subsection{Train and Test dataset}
For this study, we utilise three different populations of synthetic planetary systems. Initially, we train our algorithm on a `training set', where each system is labelled as `True' (host an ELP) or `False' (does not host an ELP). Thus, the algorithm learns to recognise which systems host an ELP and which are ELP-free. This training set comprises the majority (80\%) of our synthetic systems. Once the algorithm is trained, we can test it on a `test set', which is an unlabeled dataset on which the model makes predictions to analyse its responses and determine its precision and recall scores. The test set consists of the remaining 20\% of the dataset to ensure that the systems on which we test the algorithm are not the same as those on which it was trained, which would bias the results.\\
It is also important to ensure that the different proportions are respected in both datasets. For example, if the test set comprises 80\% of systems with an ELP while the overall proportion in the population is 40\%, the test is biased.\\
To ensure a consistent training and test set, we divided the systems with an ELP from the systems without an ELP ($\neg$ELP) in each distinct population, resulting in six subgroups (1~$M_{\odot}$/ELP, 1~$M_{\odot}$/$\neg$ELP, 0.5~$M_{\odot}$/ELP, 0.5~$M_{\odot}$/$\neg$ELP, 0.2~$M_{\odot}$/ELP, and 0.2~$M_{\odot}$/$\neg$ELP, where ELP means the systems with at least one ELP and $\neg$ELP means the systems without any ELPs). Then, 80\% of each subgroup constitutes the training set, and the remaining 20\% forms the test set.When creating training and test sets with the three mixed populations, we ensure that the proportion of each population remains the same. Thus, we choose the population with the fewest systems and adjust the other populations to match this number. This way, we ensure that we have the same proportion of systems from each population (1, 0.5, and 0.2 $M_{\odot}$) in both datasets. On the other hand, we do not scale the number of systems with and without ELP. The proportion of systems with ELP in each population is a feature in itself that the model must account for.\\
Once trained on the training set, we test the algorithm on the test set and calculate its performance using the different scores. We then apply it to a list of observed planetary systems to predict whether a system is likely to host an ELP or not. This likelihood is characterised by the algorithm's voting rate.
%________________________________________________________________________________
\section{Results} \label{sec:results}
%________________________________________________________________________________
To optimise the classification model, we first conduct several tests on the training data and the systems' descriptive features, described in the following paragraphs. Once the best strategy is identified, we use the model trained on a sample of 1567 observed systems to predict the presence of an ELP.

%-------------------------------------------------------------------------------------------------------
\subsection{Features analysis}
\begin{figure}
    \centering
    \includegraphics[width=9cm]{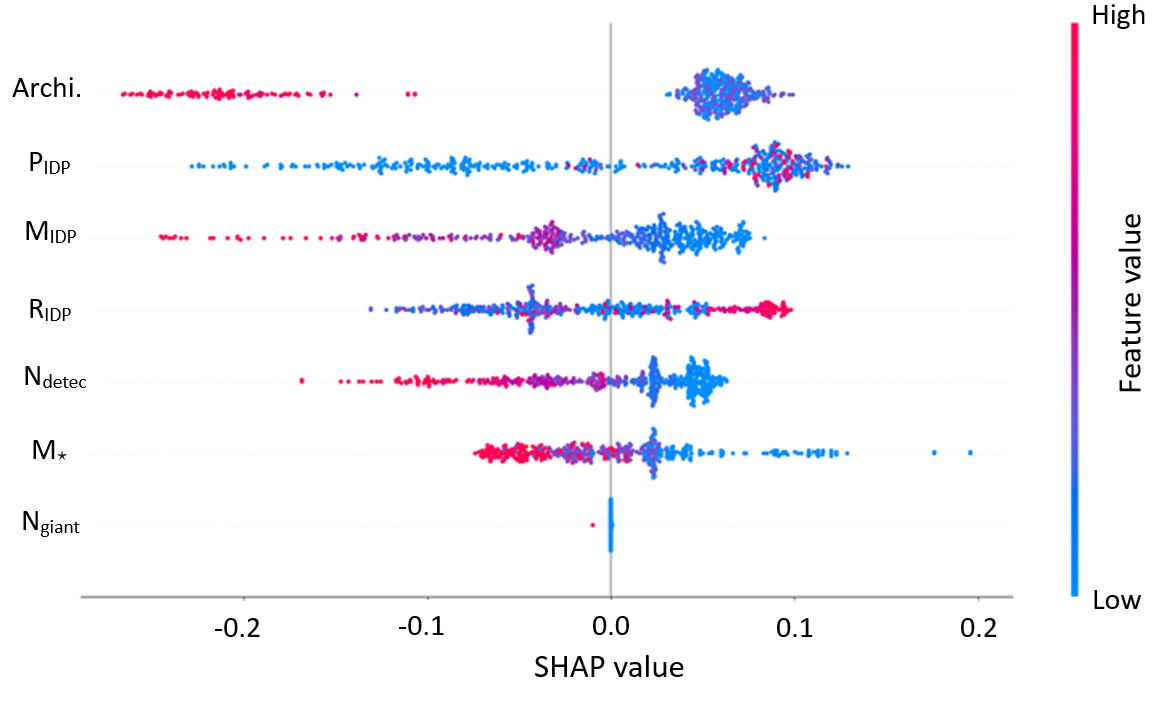}
    \caption{Bee swarm plot of the seven features considered. The x-axis represents the SHAP value of the feature for each instance, and the y-axis represents the seven features considered ranked from the most important (top) to the least (bottom). The colour of the dots represents the value of the feature itself, red being high values and blue being low values.}
    \label{fig:bee}
\end{figure}

%-------------------------------------------------------------------------------------------------------
As discussed in Sections \ref{subsubsec:featuresD24} and \ref{subsubsec:featuresmanual}, we have selected seven potentially useful descriptive features for this study. As mentioned in Section \ref{subsec:features}, we need to identify the ones that provide the most information about the presence of an Earth-like planet to maximise the performance of the Random Forest model.\\
To select the most useful features, we conduct a feature analysis. We apply the Shapley value concept to assess the features' importance of all the features described in Sect. \ref{subsec:features}. Originating from cooperative game theory, Shapley values are frequently used in machine learning to analyse the importance of features. They represent each feature's contribution to the model's prediction by evaluating all possible feature combinations and measuring the impact of adding or removing each feature.\\
Fig. \ref{fig:bee} presents a bee swarm plot where each point represents the SHAP (SHapley Additive exPlanations) \citep{Lundberg2017} value of a specific feature for an individual instance in the dataset. This visualisation shows how each feature's contribution affects the model's prediction for that instance. The y-axis lists the features from most influential (top) to least influential (bottom), while the x-axis shows the SHAP value of each feature for each dataset instance. Negative SHAP values indicate a stronger contribution to the decision `without ELP', while positive values indicate a stronger contribution to the decision `with ELP'. Additionally, the colour of the points represents the feature value itself, with higher values in red and lower values in blue. In this diagram, we have removed systems with no visible planets to facilitate readability. Indeed, when no planets are visible in a system, the IDP's mass, radius, and period are set to -1000 to indicate the absence of values for these features. This procedure results in the final bee swarm plot being polluted by very low values, making it difficult to interpret the values of the different features.\\
From Fig. \ref{fig:bee}, we observe that architecture emerges as the most important feature, with lower values indicating a greater likelihood of containing an ELP. We assigned values from 1 to 5 to the architectures (n = 1: 1, Low-mass: 2, Anti-Ordered: 3, Ordered: 4, Mixed: 5). A low value for the architecture indicates either n = 1 or Low-mass, which are the dominant classes hosting an ELP.\\
%The mass of the central star is also significant. Systems with less massive stars exhibit a higher likelihood of being classified in the 'with ELP' category \textcolor{red}{(ce qui est bizarre}. 
The period and mass of the innermost detectable planet (IDP) also play a significant role. Systems where the IDP has a greater distance from the star are more likely to be classified as `with ELP', consistent with the findings of \citetalias{Davoult2024}. Conversely, systems with a less massive IDP are also more likely to be classified as `with ELP', further supporting \citetalias{Davoult2024}'s results.\\
The impact of the IDP's radius is nuanced, as observed in \citetalias{Davoult2024}: for a given architecture, either a larger or smaller IDP radius can be more favourable for detecting an ELP. This makes it a more difficult characteristic to use, because there is no clear cut.\\
The number of visible planets, although lower ranked, also provides valuable information: the more visible planets there are, the less likely the system is to host an ELP, which confirms the observations discussed in section \ref{subsubsec:featuresmanual}.\\
However, the influence of the central star is not consistent with the first analysis, which showed that systems around stars of 0.2~$M_{\odot}$ had proportionately fewer ELPs (only 40\% of systems, compared with 75\% and 60\% for 0.5 and 1~$M_{\odot}$ respectively). In this representation, we have removed systems without planets larger than 0.5~$M_{\oplus}$ (see Appendix A of \citetalias{Davoult2024}). However, the vast majority of these empty systems are systems without any ELP, which reverses the proportion of systems with ELP if they are not considered. In this representation, we therefore have systems with a low-mass star (blue represents stars of 0.2~$M_{\odot}$) classified as having an ELP, systems with a star of 1~$M_{\odot}$ (red) classified as having no ELP, and systems with a star of 0.5~$M_{\odot}$, being hard to classify because they have almost equal numbers of systems with and without ELP.\\
Finally, the number of giant planets in the system provides limited information. Specifically, as the number of giant planets increases, the model tends to classify the system as ELP-free. However, in the absence of giant planets, the model has difficulty making a clear decision. This mirrors the data observed: systems with giant planets are much less likely to have an ELP, while those without giant planets may or may not host an ELP.\\
%\textcolor{red}{This is likely due to a common bias in Random Forest models, where features with a large number of distinct values tend to be overvalued. The algorithm calculates feature importance based on the reduction in uncertainty. However, when a feature has many unique values, the reduction in uncertainty is artificially inflated because the model can split more frequently based on that feature.} \textcolor{blue}{Mais Nvis est aussi petite values, et archi aussi}
%\textcolor{red}{C'est notamment le cas de l'architecture qui ne peut prendre que les valeurs 0, 1, 2, 3, 4 et 5, la masse de l'étoile qui ne peut prendre que les valeurs 0.2, 0.5 et 1, le nombre de planètes géantes (des valeurs discrètes de 0 à 3 \textcolor{blue}{(à vérifier)}) et le nombre de planète visible (des valeurs discrètes également de 0 à ?? \textcolor{blue}{(à vérifier)}.}\\
Based on Fig. \ref{fig:bee}, it appears that the most important features are the architecture, and the mass and period of the innermost detectable planet (IDP).\\
To compare the performance of the Random Forest Classifier based on the descriptive features used, we conducted four tests, each time changing the descriptive features. The first test includes all features, the second includes only manually selected features (Sect. \ref{subsubsec:featuresmanual}), the third includes features derived from \citetalias{Davoult2024} (Sect. \ref{subsubsec:featuresD24}), and the fourth includes the top features selected from Fig. \ref{fig:bee}. Those four tests are resumed in Table \ref{tab:test_features}.\\
\begin{table}[]
        \caption[]{List of planetary systems features used in each test.}
        \begin{tabular}{|l|l|}
        \hline
        N° test & Features included\\
        \hline
        Test n°1 & all\\
        Test n°2 & $N_{\text{detec}}$, $N_{\text{giant}}$, $M_{\text{IDP}}$, $M_{\star}$\\
        Test n°3 & $R_{\text{IDP}}$, $M_{\text{IDP}}$, $P_{\text{IDP}}$, architecture, $M_{\star}$\\
        Test n°4 & $M_{\text{IDP}}$, $P_{\text{IDP}}$, architecture, $M_{\star}$\\
        \hline
    \end{tabular}
    \label{tab:test_features}
\end{table}
Table \ref{tab:results_testfeatures} displays the Random Forest results for voting rate thresholds of 50\% (default), 70\%, 80\%, and 90\%. For each test and each threshold, the table shows the confusion matrix and the precision score (PS). As a reminder, a confusion matrix is constructed as 
$\begin{pmatrix}
    TN & FP\\
    FN & TP 
\end{pmatrix}$, with TN and FN representing True Negatives and False Negatives, and TP and FP representing True Positives and False Positives. The confusion matrix is beneficial in unbalanced datasets like this one. It allows us to assess not only whether the model correctly classifies the instances but also its performance (very few false positives and true positives indicate that the model struggles to understand what a system with an ELP looks like).\\
Unsurprisingly, for the default threshold of 50\%, the results are fair but not excellent. True positive answers account for just above 80\% of all positive answers. As the threshold increases, the precision score improves, indicating the model's ability to recognise patterns that distinguish systems with ELPs. Increasing the precision score also means increasing the TP/FP ratio. However, in these cases, we also notice an increase in false negatives (FN), indicating that the model becomes more conservative, missing more positives in its effort to reduce false positives. From a threshold of 80\%, all three tests show precision scores above 0.9, indicating that true positives account for 90\% of the model's positive predictions, demonstrating its excellent capability.\\
Although the four tests show similar overall performance, a closer look into the confusion matrix reveals that Test n°1 is less effective than the other three tests. Specifically, above 90\%, Test n°1 shows fewer TP and FP, resulting in fewer overall positive answers. While the ratio between TP and FP remains similar, the lower number of total positive answers indicates that it recognises systems with an ELP less effectively. 
Tests n°2 exhibits a slightly lower PS for all thresholds above 80\%. Test~n°3 and Test~n°4 show same precision score for thresholds above 80\% but Test~n°4 exhibit slightly fewer False Negative and more True Positive, this reflects its ability to recognise a system with an ELP more effectively. For this reason, Test~N°4 is used in the remainder of the study.\\
\begin{table*}[]
        \caption[]{Performance results of the model trained on four different tests. The Test n°1 uses all the features.}
        \begin{tabular}{|c|c|c|c|c|}
        \hline
        & Test n°1 & Test n°2 & Test n°3 & Test n°4\\
        \hline
        50\% & 
        $\begin{pmatrix}
            1035 & 285\\
            148 & 1451
        \end{pmatrix}$ PS=0.84 & $\begin{pmatrix}
            1027 & 331\\
            131 & 1430
        \end{pmatrix}$ PS=0.81 & $\begin{pmatrix}
            990 & 308\\
            172 & 1449
        \end{pmatrix}$ PS=0.83 & $\begin{pmatrix}
            1061 & 287\\
            144 & 1427
        \end{pmatrix}$ PS=0.83\\
         \hline
        70\% & 
        $\begin{pmatrix}
            1095 & 225\\
            208 & 1391
        \end{pmatrix}$ PS=0.86 & $\begin{pmatrix}
            1100 & 258\\
            215 & 1346
        \end{pmatrix}$ PS=0.84 & $\begin{pmatrix}
            1063 & 285\\
            226 & 1395
        \end{pmatrix}$ PS=0.86 & $\begin{pmatrix}
            1106 & 242\\
            194 & 1377
        \end{pmatrix}$ PS=0.85 \\
         \hline
        80\% & 
        $\begin{pmatrix}
            1309 & 11\\
            859 & 740
        \end{pmatrix}$ PS=0.99 & $\begin{pmatrix}
            1337 & 21\\
            811 & 750
        \end{pmatrix}$ PS=0.97 & $\begin{pmatrix}
            1281 & 17\\
            860 & 760
        \end{pmatrix}$ PS=0.98 & $\begin{pmatrix}
            1328 & 20\\
            788 & 783
        \end{pmatrix}$ PS=0.98\\
        \hline
        90\% & 
        $\begin{pmatrix}
            1316 & 4\\
            1423 & 176
        \end{pmatrix}$ PS=0.98 & $\begin{pmatrix}
            1345 & 13\\
            973 & 588
        \end{pmatrix}$ PS=0.98 & $\begin{pmatrix}
            1290 & 8\\
            942 & 679
        \end{pmatrix}$ PS=0.99 & $\begin{pmatrix}
            1343 & 5\\
            861 & 710
        \end{pmatrix}$ PS=0.99\\
         \hline
    \end{tabular}
    \label{tab:results_testfeatures}
    \tablefoot{Test n°2 uses the number of detectable planets, number of giant planets, mass of the innermost detectable planet and central star's mass. Test n°3 uses the radius, mass and period of the innermost detectable planet, the architecture of the system and the mass of the central star. Test n°4 corresponds to the most important features: the mass and period of the innermost detectable planet, the architecture of the system and the mass of the central star. The different percentages correspond to different voting rate thresholds of the Random Forest.}
    \end{table*}

%--------------------------------------------------------------------------------
\subsection{Population analysis}
%--------------------------------------------------------------------------------
Now that we have determined the features to use, we need to decide on which populations of synthetic systems the model should be trained to achieve optimal performance. Several strategies are considered:
\begin{itemize}
    \item Mass-Specific Training: To predict the outcome of a system, we use a model trained exclusively on systems with similar central star's mass. The star's mass is not a feature provided to the model but is considered when choosing which training data to use.\\
    \item Global Population Training: We train the model on a combined population that includes systems with central stars of different masses, regardless the central star's mass. Here, the star's mass is an input to the model so that it can differentiate between different types of systems mixed in the overall training population.\\
    \item Subset Training: We create subsets of training data. For example, we train the model on populations of systems with 1 and 0.5~$M_{\odot}$ stars together because they share similarities, particularly in terms of the proportion of systems with an ELP (60\% for solar-mass stars and 74\% for 0.5~$M_{\odot}$ stars, compared to only 40\% for 0.2~$M_{\odot}$ stars), and separately on the population with 0.2~$M_{\odot}$ stars.
\end{itemize}
We construct several training populations to evaluate which strategy is best. These populations are summarised in Table \ref{tab:list_pop}.
\begin{table}[]
        \caption[]{Different training population used in the tests.}
        \begin{tabular}{|l|l|}
        \hline
        Population name & Synthetic population included\\
        \hline
        MS-1 & 1~$M_{\odot}$\\
        MS-0.5 & 0.5~$M_{\odot}$\\
        MS-0.2 & 0.2~$M_{\odot}$\\
        Subset & 1~$M_{\odot}$ + 0.5~$M_{\odot}$\\
        Global & 1~$M_{\odot}$ + 0.5~$M_{\odot}$ + 0.2~$M_{\odot}$\\
        \hline
    \end{tabular}
    \label{tab:list_pop}
    \end{table}
We then train the Random Forest, and test it for each strategy. The results are shown in Table \ref{tab:results_testpop}.\\
%We want to notify the reader that this table is read in the opposite direction compared to the previous table; columns represent different thresholds.
A quick glance shows that the last two populations (Subset and Global) yield very similar results.
The models trained on the three populations built for the Mass-Specific strategy (MS-1, MS-0.5 and MS-0.2) show however different results: the model trained on MS-1 and MS-0.5 have better results than on MS-0.2. This can be explained both by the fact that the population of 0.2~$M_{\odot}$ stars is unbalanced negatively (only 40\%) of systems host an ELP) but also because there are a lot less systems in this population than in the two others. After correction for empty systems, the MS-0.2 has 4862 systems, MS-0.5 has 10158 systems and MS-1 has 20365 systems. Although above a threshold of 90\%, all populations yield the same result (PS = 0.99) except for MS-0.2 (PS = 0.94), we chose the mass-specific strategy. This strategy allows for the maximum use of training data and helps avoid overfitting. For the subset training and global training strategies, populations are scaled to have the same number of systems. In other words, systems are randomly removed from the larger populations to match the number of data points in the smallest population.\\
\begin{table*}[]
        \caption[]{Performance results of the model trained on different populations.}
        \begin{tabular}{|c|c|c|c|c|}
        \hline
        & 50\% & 70\% & 80\% & 90\%  \\
        \hline
        MS-1 & 
        $\begin{pmatrix}
            824 & 205\\
            238 & 2807
        \end{pmatrix}$ PS=0.93 & $\begin{pmatrix}
            945 & 84\\
            361 & 2684
        \end{pmatrix}$ PS=0.97 & $\begin{pmatrix}
            967 & 62\\
            406 & 2639
        \end{pmatrix}$ PS=0.98 & $\begin{pmatrix}
            1011 & 18\\
            544 & 2501
        \end{pmatrix}$ PS=0.99\\
         \hline
        MS-0.5 & 
        $\begin{pmatrix}
            23 & 466\\
            28 & 1516
        \end{pmatrix}$ PS=0.77 & $\begin{pmatrix}
            43 & 446\\
            69 & 1475
        \end{pmatrix}$ PS=0.77 & $\begin{pmatrix}
            482 & 7\\
            1318 & 226
        \end{pmatrix}$ PS=0.97 & $\begin{pmatrix}
            487 & 2\\
            1364 & 180
        \end{pmatrix}$ PS=0.99 \\
         \hline
        MS-0.2 & 
        $\begin{pmatrix}
            837 & 3\\
            91 & 42
        \end{pmatrix}$ PS=0.93 & $\begin{pmatrix}
            838 & 2\\
            93 & 40
        \end{pmatrix}$ PS=0.95 & $\begin{pmatrix}
            838 & 2\\
            95 & 38
        \end{pmatrix}$ PS=0.95 & $\begin{pmatrix}
            839 & 1\\
            117 & 16
        \end{pmatrix}$ PS=0.94 \\
        \hline
        Subset & 
        $\begin{pmatrix}
            437 & 566\\
            140 & 2923
        \end{pmatrix}$ PS=0.84 & $\begin{pmatrix}
            502 & 501\\
            209 & 2854
        \end{pmatrix}$ PS=0.85 & $\begin{pmatrix}
            964 & 39\\
            1556 & 1507
        \end{pmatrix}$ PS=0.98 & $\begin{pmatrix}
            982 & 21\\
            1672 & 1391
        \end{pmatrix}$ PS=99 \\
         \hline
        Global & 
        $\begin{pmatrix}
            1061 & 287\\
            144 & 1427
        \end{pmatrix}$ PS=0.83 & $\begin{pmatrix}
            1106 & 242\\
            194 & 1377
        \end{pmatrix}$ PS=0.85 & $\begin{pmatrix}
            1328 & 20\\
            788 & 783
        \end{pmatrix}$ PS=0.98 & $\begin{pmatrix}
            1341 & 5\\
            861 & 710
        \end{pmatrix}$ PS=99 \\
         \hline
    \end{tabular}
    \label{tab:results_testpop}
    \tablefoot{MS-1, MS-0.5 and MS-0.2 are used in the Mass-Specific training strategy and correspond to individual populations. Subset corresponds to the population of 0.5 and 1~$M_{\odot}$ stars combined for the Subset strategy. Global population is the three populations combined for the global strategy.}
    \end{table*}

%---------------------------------------------------------------------------------------------------------------
\subsection{Prediction of detection} \label{prediction}
%---------------------------------------------------------------------------------------------------------------
The developed and trained model can now be used to predict which systems are most likely to host an ELP. We use a sample of 1567 known systems around MKG stars from \texttt{exoplanet.eu}\footnote{available at https://exoplanet.eu/catalog/} \citep{Schneider2011} in which at least one planet and its mass is known, regardless its detection method.\\
The dataset is then divided into three subsets: 1025 systems with central star masses between 0.7 and 1.2 $M_{\odot}$, 342 systems with central star masses between 0.35 and 0.7 $M_{\odot}$, and 200 systems with central star masses less than 0.35 $M_{\odot}$. Each subset corresponds to a specific training dataset: 1, 0.5, and 0.2 $M_{\odot}$, respectively. We apply the same observational bias to each subgroup as described in section \ref{sec:bias}, according to the mass of the central star. For each system, we extract the corresponding features for Test n°4: the system's architecture with planets that overpass the observational bias, the mass, and the period of the IDP, if one planet remained in the system after applying the bias. We then use the model trained on the populations corresponding to each subset to obtain the voting rate of each planetary system.\\
Among the 1567 total systems in the three subsets, 51 achieved a voting rate of more than 90\%. We exclude binary systems because the Bern model produced only single stars and the habitable zone is calculated differently in binaries \citep{Haghighipour2015}, and the 44 remaining systems with their associated voting rates are listed in Table \ref{tab:result_syst}.\\

\begin{table}[]
        \caption[]{List of 44 systems achieving a voting rate (VR) of over 90\%.}
        \begin{tabular}{|l|l|l|}
        \hline
        System & VR & Reference\\
        \hline
        \hline
        \multicolumn{3}{|c|}{G stars}\\
        \hline
        HD 103949 & 96\% & {\footnotesize \cite{Feng2019}}\\
        HD 42618 & 95\% & {\footnotesize \cite{Fulton2016}}\\
        HD 85390 & 93\% & {\footnotesize \cite{Wittenmyer2019}}\\
        HIP 41378 & 97\% & {\footnotesize \cite{Vanderburg2016}}\\
        Kepler-22 & 92\% & {\footnotesize \cite{Borucki2012}}\\
        Kepler-538 & 96\% & {\footnotesize \cite{Mayo2019}}\\
        {\footnotesize KMT-2021-BLG-0171L} & 94\% & {\footnotesize \cite{Yang2022}}\\
        \hline
        \hline
        \multicolumn{3}{|c|}{late-K and early-M stars}\\
        \hline
        {\footnotesize OGLE-2017-BLG-1691L} & 97\% & {\footnotesize \cite{Han2022}}\\
        GJ 685 & 92\% & {\footnotesize \cite{Pinamonti2019}}\\
        Gl 514 & 95\% & {\footnotesize \cite{Damasso2022}}\\
        HD 147379 & 96\% & {\footnotesize \cite{Reiners2018}}\\
        HD 211970 & 92\% & {\footnotesize \cite{Feng2019}}\\
        HIP 71135 & 95\% & {\footnotesize \cite{Feng2019}}\\
        K2-286 & 94\% & {\footnotesize \cite{Diez2019}}\\
        {\footnotesize KMT-2022-BLG-0440L} & 98\% & {\footnotesize \cite{Zhang2023}}\\
        {\footnotesize KMT-2022-BLG-0475} & 92\% & {\footnotesize \cite{Han2023}} \\
        {\footnotesize OGLE-2007-BLG-368L} & 96\% & {\footnotesize \cite{Sumi2010}}\\
        {\footnotesize OGLE-2015-BLG-0966L} & 96\% & {\footnotesize \cite{Street2016}}\\
        {\footnotesize OGLE-2015-BLG-1670L} &  96\% & {\footnotesize \cite{Ranc2019}}\\
        {\footnotesize OGLE-2018-BLG-0506} & 96\% & {\footnotesize \cite{Hwang2022}}\\
        {\footnotesize OGLE-2018-BLG-0516} & 95\% & {\footnotesize \cite{Hwang2022}}\\
        {\footnotesize OGLE-2018-BLG-1126} & 96\% & {\footnotesize \cite{Gould2022}}\\
        {\footnotesize OGLE-2018-BLG-1185} & 96\% & {\footnotesize \cite{Kondo2021}}\\
        {\footnotesize TCP J050742+244755} & 97\% & {\footnotesize \cite{Nucita2018}}\\
        TOI-1231 & 91\% & {\footnotesize \cite{Burt2021}}\\
        TOI-2285 & 91\% & {\footnotesize \cite{Fukuti2022}}\\
        \hline
        \hline
        \multicolumn{3}{|c|}{late-M stars}\\
        \hline
        G 9-40 & 91\% & {\footnotesize \cite{Stefansson2020}}\\
        GJ 1061 & 98\% & {\footnotesize \cite{Dreizler2020}}\\
        GJ 1132 & 94\% & {\footnotesize \cite{Bonfils2018}}\\
        GJ 273 & 92\% & {\footnotesize \cite{Pozuelos2020}}\\
        GJ 3323 & 95\% & {\footnotesize \cite{Astudillo2017}}\\
        GJ 357 & 98\% & {\footnotesize \cite{Jenkins2019}}\\
        GJ 3929 & 91\% & {\footnotesize \cite{Kemmer2022}}\\
        GJ 3988 & 91\% & {\footnotesize \cite{Gorrini2023}}\\
        GJ 581 & 98\% & {\footnotesize \cite{Stauffenberg2024}}\\
        L 98-59 & 98\% & {\footnotesize \cite{Demangeon2021}}\\
        LHS 1140 & 95\% & {\footnotesize \cite{Dittman2017}}\\
        Teegarden's & 95\% & {\footnotesize \cite{Dreizler2024}}\\
        TOI-1680 & 91\% & {\footnotesize \cite{Ghachoui2023}}\\
        TOI-2096 & 98\% & {\footnotesize \cite{Pozuelos2023}}\\
        TOI-2136 & 91\% & {\footnotesize \cite{Gan2022}}\\
        TOI-237 & 92\% & {\footnotesize \cite{Waalkes2021}}\\
        Wolf 1061 & 98\% & {\footnotesize \cite{Astudillo2017}}\\
        YZ Cet & 98\% & {\footnotesize \cite{Astudillo2017b}}\\
        \hline
    \end{tabular}
    \label{tab:result_syst}
    \end{table}

\begin{figure*}
    \centering
    \includegraphics[width=15cm]{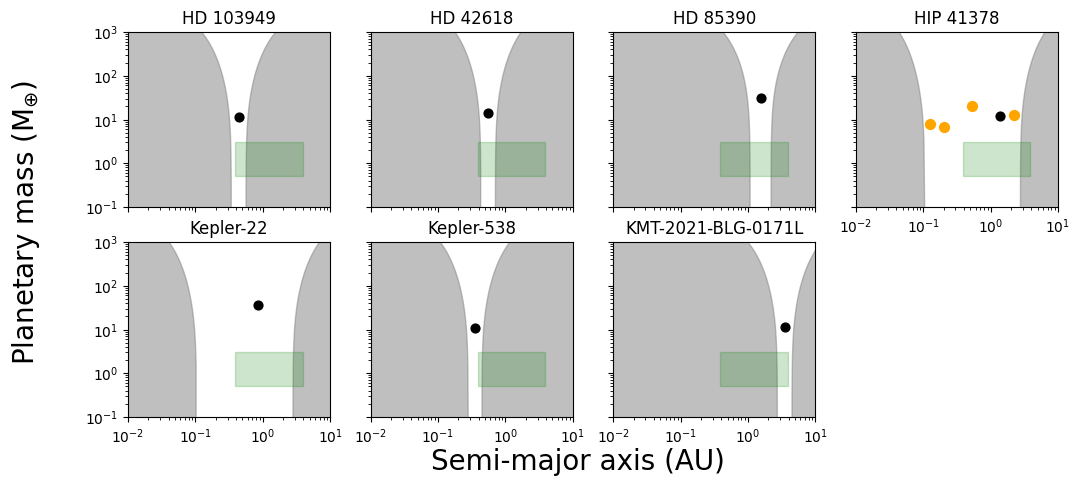}
    \caption{Systems around G stars with a resulting voting rate above 90\%. The green areas represent the definition of an Earth-like planet in the study in terms of equilibrium temperature and mass. The grey areas represent the combinations of mass and semi-major axis for which the Hill-stability criterion is met with the addition of a new planet. The black dots correspond to planets for which we know the mass, and the orange dots correspond to planet for which we only know the radius, and the mass has been derived thanks to the work of \cite{Parc2024}.}
    \label{fig:result_multi_1Msun}
\end{figure*}

\begin{figure*}
    \centering
    \includegraphics[width=16cm]{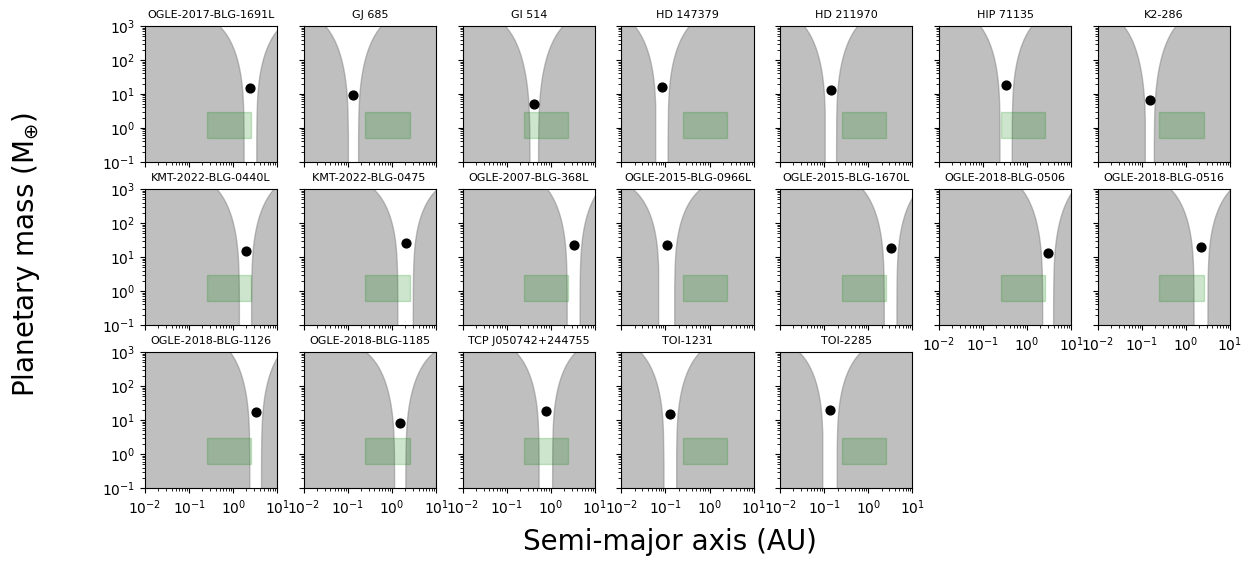}
    \caption{Systems around early-M and late-K stars with a resulting voting rate above 90\%. The green areas represent the definition of an Earth-like planet in the study in terms of equilibrium temperature and mass. The grey areas represent the combinations of mass and semi-major axis for which the Hill-stability criterion is met with the addition of a new planet. The black dots correspond to the planets already known in these systems.}
    \label{fig:result_multi_05Msun}
\end{figure*}

\begin{figure*}
    \centering
    \includegraphics[width=16cm]{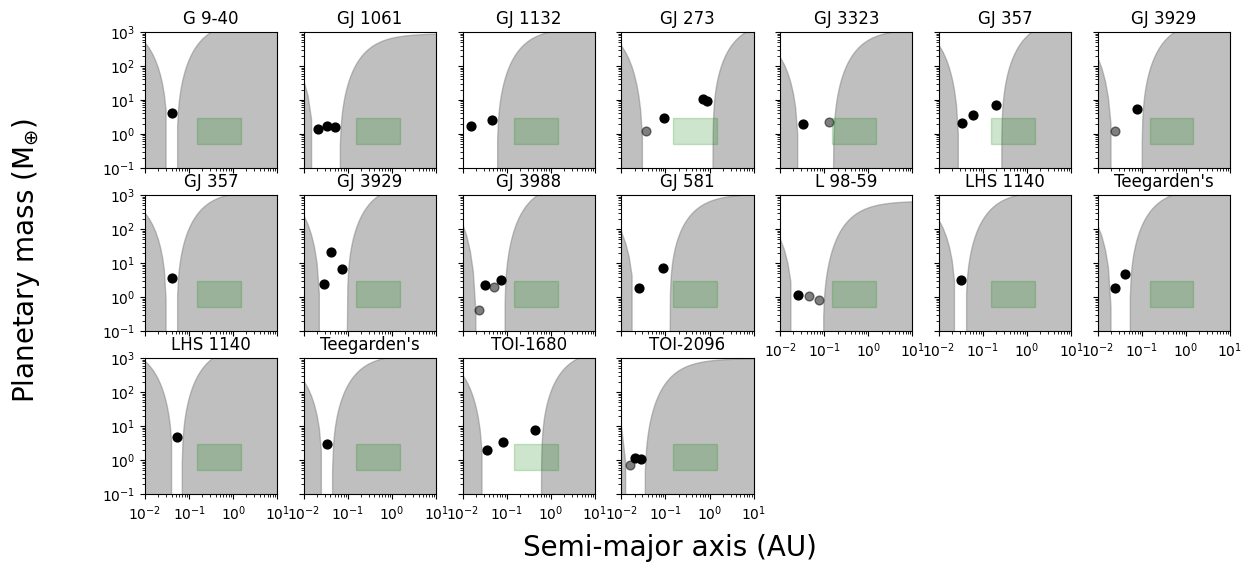}
    \caption{Systems around late-M stars with a resulting voting rate above 90\%. The green areas represent the definition of an Earth-like planet in the study in terms of equilibrium temperature and mass. The grey areas represent the combinations of mass and semi-major axis for which the Hill-stability criterion is met with the addition of a new planet. The dots represent the planets already known in those systems: the black dots for planets with a RV semi-amplitude above the threshold of detection bias (detectable planets) and the grey dots for the planets with a RV semi-amplitude below this threshold. Only the detectable planets count in the calculation of the architecture of the systems.}
    \label{fig:result_multi_02Msun}
\end{figure*}

To evaluate the possibility of a planet's existence in these systems, we use the stability criterion from \cite{Fabrycky2014}, which was also employed in \cite{Chen2024}. The Hill-stability criterion $H$ is defined as: 
\begin{equation}
    H = \frac{a_{\text{out}}-a_{\text{in}}}{R_{\text{H}}},
\end{equation}
 with $a_{\text{in}}$ and $a_{\text{out}}$ referring to the semi-major axes of the inner and outer planets, respectively, and $R_{\text{H}}$ being the mutual Hill radius relevant for dynamical interactions \citep{Fabrycky2014}:
\begin{equation}
    R_{\text{H}}=\left(\frac{M_{\text{in}}+M_{\text{out}}}{M_{\star}}\right)^{1/3}\frac{(a_{\text{in}}+a_{\text{out}})}{2}, 
\end{equation}
with $M_{\text{in}}$ and $M_{\text{out}}$ being the masses of the inner and outer planets, respectively, and $M_{\star}$ being the mass of the star.
For a two-planet system, \cite{Chen2024} defines $H$ > 7.1 for the system to be stable. In a system with more than two planets, a more stringent criterion is used:
\begin{equation}
    H_{\text{in}}+H_{\text{out}} > 18,
\end{equation}
with $H_{\text{in}}$ and $H_{\text{out}}$ being the Hill-stability criteria for the inner and outer planet pairs.\\
Figure \ref{fig:result_multi_1Msun}, \ref{fig:result_multi_05Msun} and \ref{fig:result_multi_02Msun} present the systems identified as potential candidates for hosting an ELP in a mass-semi-major axis diagram around G stars, early-M and late-K stars, and late-M stars respectively. The dots indicate the already existing planets: black dots represent visible planets for which we know the mass and are the planets used to assess the voting rates of a system. Grey dots represent planets for which we know the mass but with a RV semi-amplitude lower than the detection bias (they do not contribute to the determination of the architecture). Finally, orange points represent planets for which we only know the radius, and the mass has been derived using the mass-radius relationship from \cite{Parc2024}. The latters are only used to evaluate the stability of a system with an additional planet, but they are not used in the model. The green areas outline the regions defining an ELP in terms of mass and equilibrium temperature. The grey areas correspond to regions where the Hill-stability criterion is met and where the presence of an additional planet is possible. The green and grey areas overlap in most of systems identified by our algorithm, indicating the potential for an ELP in these systems. Particularly, for G stars, only the system HIP 41378 does not seem stable with the addition of an ELP. However, if we knew the precise mass of the four planets represented by an orange dot, the model would not have classified it in the category `with ELP'. For late-K and early-M stars, all of the systems seem stable while adding an ELP. Finally, for late-M stars, only GJ 273 does not seem stable while adding an ELP. These results highlight the effectiveness of our model: 95.5\% of the systems identified as likely to host an ELP can theoretically host one.

\section{Discussion and conclusion} \label{sec:conclusion}
\subsection{Discussion}
The model presented in this work presents a few limitations and avenues to improvement that we would like to discuss.
Firstly, using this model on a sample of known and observed planetary systems involves assuming that the Bern model accurately replicates observed planetary systems and that the properties correlated with the presence of an Earth-like planet in synthetic systems are the same in real planetary systems. In reality, the synthetic systems modelled with the Bern model only partially resemble actual planetary systems. Other studies \citep{Mulders2019,NGPPS3,NGPPS4,NGPPS6,Mishra2023,Burn2024, Emsenhuber_inprep} have shown that populations of planetary systems replicate the basic patterns observed in actual planetary populations. The populations of synthetic systems calculated using the Bern model demonstrate a positive correlation between the occurrence of inner Super-Earths and cold Giants \citep{NGPPS3}, albeit weaker than those observed in studies such as \cite{Zhu&Wu2018}. Additionally, the model captures trends related to dependencies on stellar metallicity \citep{NGPPS3,Emsenhuber_inprep} and stellar mass \citep{NGPPS4}, as well as patterns in period ratio distributions \citep{Mulders2019,NGPPS4, Emsenhuber_inprep} and eccentricity distributions \citep{NGPPS4,Emsenhuber_inprep}. Notable architectural features include similarities and mass or size ordering \citep{NGPPS6}, the `peas-in-a-pod' structure \citep{NGPPS6}, a bimodal mass function distinguishing sub-Neptunes and Gas Giants \citep{Mulders2019,Emsenhuber_inprep}, and a mean observed multiplicity of approximately 1.6 \citep{Emsenhuber_inprep}.
Despite these successes, the model has limitations in reproducing certain observed characteristics of planetary populations. For example, the positive correlation between Super Earths and cold Giants is weaker than observed \citep{NGPPS3}. Moreover, the model predicts an overproduction of planets per system—by at least a factor of 1.7 \citep{Mulders2019,Emsenhuber_inprep}. Synthetic planets also tend to be closer to their stars than observed \citep{Mulders2019,Emsenhuber_inprep}, and the mass distribution does not align precisely with observations \citep{Emsenhuber_inprep}. Finally, the model produces an excess of planets in or near mean-motion resonances \citep{Mulders2019,NGPPS4,Emsenhuber_inprep}, which is inconsistent with the distribution seen in observed systems. In summary, planetary system populations are realistic at the system and architectural scale rather than at the individual planet scale. These results suggest that we can consider synthetic planetary systems as proxies for real systems when examining architectures and correlations between planetary properties.\\
Another limitation of the model is the limited amount of training data: there are only between 5000 to 20000 instances depending on the populations, due to the time required to generate synthetic systems. This issue could be addressed by the upcoming work of \cite{Alibert_inprep}, which employs a transformer-based generative model to emulate the Bern model and generate millions of synthetic planetary systems in an hour.\\
Finally, another weakness of the study is the simplistic approach to handling observational bias, which does not account for various factors that may influence planet detection, such as stellar activity, the presence of other planets in the system, observation frequency, and orbital period. Forthcoming works should address this issue.% which models radial velocity signals of synthetic planetary systems from the Bern model considering stellar activity, observation strategy, and variables such as stellar distance and system inclination, promises a more realistic treatment of the radial velocity bias.\\

\subsection{Conclusion}
In this work, we have developed a model using a Random Forest Classifier to predict which known planetary systems are most likely to host an Earth-like planet. The model was trained on a dataset of synthetic planetary systems from the Bern model to which we applied an observational bias to extract observable properties. We conducted tests to determine the optimal descriptive features of synthetic systems to enhance model performance, finding that the mass, period of the innermost detectable planet (IDP), and system architecture are the three properties that provide the most information about the presence of an Earth-like planet. These findings are consistent with the results of \cite{Davoult2024}.\\
The model demonstrated excellent performance, achieving a precision score of up to 0.99 on the test datasets, which means that 99\% of the positive predictions were True Positives. This result proves that the model can accurately identify the properties of systems with and without ELPs within a dataset derived from the Bern model.\\
Therefore, we used the model to predict the presence of an Earth-like planet in a sample of 1567 observed GKM systems, for which we know at least one planet and the properties necessary for the model to function (the mass and semi-major axis or period of at least one planet and the mass of the central star). The results indicate that 44 systems (listed in Table \ref{tab:result_syst}) exhibit architectures suggesting the presence of an Earth-like planet. Further study of the stability state of these systems with the addition of a new planet has shown that 95.5\% of those systems would remain stable with the addition of an Earth-like planet.\\
We caution that the results heavily rely on the Bern model and should be interpreted cautiously. However, we recommend prioritising the study of these systems because both positive and negative outcomes provide conclusive findings. Negative results would indicate that the Bern model is having difficulty reproducing the architecture of the systems, and would be a path for improvement. In the context of predicting exoplanet detection using global models of planetary formation, it is crucial to link observations with theoretical models to rigorously test them closely.
%Enfin, un des désavantage des Random Forest est la différence de traitement entre les features with a large number of distinct values et les features avec seulement quelques valeurs distinctes. En effet, lorsqu'un feature a beaucoupde valeurs distincte, un decision tree peut spliter plus fréquement et l'importance de cette feature est overevaluated. Dans notre cas

\begin{acknowledgements}
This work has been carried out within the framework of the National Centre of Competence in Research PlanetS supported by the Swiss National Science Foundation under grants 51NF40\_182901 and 51NF40\_205606. The authors acknowledge the financial support of the SNSF.
\end{acknowledgements}

\bibliographystyle{aa}
\bibliography{classifier.bib}

\end{document}